\documentclass[11pt]{article}
\usepackage[margin=1in]{geometry}
\usepackage{amsmath,amssymb}
\usepackage{booktabs}
\usepackage{hyperref}
\usepackage{natbib}
\title{Socioeconomic Inference in LLM Medical Triage:\\Same Symptoms, Different ZIP Code}
\author{Qi Han Wong\\
\texttt{wongqihan@gmail.com}\\
\url{https://github.com/wongqihan/triagebench}}
\date{June 2026}

\begin{document}
\maketitle

\begin{abstract}
We investigate whether large language models alter medical triage recommendations for identical symptoms when only the patient's socioeconomic status (SES) varies. Using three deployment-tier models (Gemini 3.5 Flash, Claude Sonnet 4.6, GPT-5.4-mini), we hold a single neurological symptom profile fixed and vary the SES signal along two channels: \textbf{explicit} (insurance status, occupation, housing) and \textbf{implicit} (a US ZIP code, with no other socioeconomic information). All three models raise their emergency-room (ER) referral rate for lower-SES patients given the explicit signal (spreads of 13--50 percentage points). The effect is in the \textbf{protective} direction: lower-SES patients are sent to the ER \emph{more} often, not less. The model's stated reasoning stays clinically near-identical across conditions, so the shift is invisible to a reasoning-trace audit. Critically, sensitivity to the \emph{implicit} ZIP-code signal is model-dependent: Gemini infers SES from geography alone, shifting its ER rate by a pooled 11.4 points across six US ZIP-code pairs ($p = 1.4 \times 10^{-7}$, same direction in 6/6 pairs), while Claude Sonnet 4.6 stays flat ($-0.1$ points) and GPT-5.4-mini shows only a small difference that is not sign-consistent (2.0 points, predicted direction in just 2 of 6 pairs)---neither a reliable ZIP-code effect, despite both responding to the explicit signal. This reveals an \textbf{explicitness gradient} in the signal: every model acts on socioeconomic status when it is stated outright, but only Gemini Flash acts on it when it must be inferred from a proxy as thin as five digits. We read this as a model-specific difference rather than a size or cost effect. A single-sentence system-prompt instruction reduces but does not eliminate the effect (Gemini's gap between low- and high-income ZIPs falls from 11.4 to 5.8 points). We release all code, prompts, and raw results.
\end{abstract}

\section{Introduction}

Large language models are increasingly deployed in symptom checkers and triage assistants \citep{singhal2023, nori2023}. A basic requirement for such systems is \textbf{socioeconomic parity in clinical assessment}: a patient's income, insurance status, or neighborhood does not change the medical danger of their symptoms, so a presentation that warrants urgent evaluation for one patient warrants it for any patient who presents the same way. A triage assistant should not let an inferred socioeconomic signal covertly shift that assessment. (Whether real-world access constraints can justify a different \emph{disposition}---routing a low-access patient to the ER because outpatient follow-up is unlikely---is a separate question we take up in Section~\ref{sec:justified}.)

We test this requirement directly. Building on prior findings that LLM triage is sensitive to gender \citep{wong2026gender} and to language-driven geographic inference \citep{wong2026lang}, we hold a single neurological symptom profile constant and vary only the patient's socioeconomic signal. We test two channels. The \textbf{explicit} channel states insurance, occupation, or housing. The \textbf{implicit} channel supplies only a US ZIP code---a five-digit proxy that carries strong socioeconomic information but states none of it. Our primary outcome is the ER referral rate.

\paragraph{Contributions.}
\begin{enumerate}
\item We show that three deployment-tier LLMs raise ER referral rates for lower-SES patients presenting identical symptoms, an effect of 13--50 percentage points under an explicit SES signal.
\item We identify the direction as \textbf{protective rather than dismissive}: lower-SES patients are routed to the ER \emph{more} often, not less.
\item We identify an \textbf{explicitness gradient}: every model acts on socioeconomic status when it is stated outright, but only one of the three (Gemini 3.5 Flash) acts on it when it must be inferred from a ZIP code (6/6 pairs, $p < 10^{-6}$). This is a model-specific difference, not one that tracks model size or price.
\item We show the bias is largely \textbf{invisible in the model's stated reasoning}: not one of 2{,}100 ZIP-conditioned Gemini responses cites income, cost, or location, so the disparity is undetectable by reasoning-trace auditing alone.
\item We release the complete experiment harness, prompts, and raw results.
\end{enumerate}

\section{Related Work}

\paragraph{Socioeconomic bias in clinical care.} Disparities in emergency triage and wait times by demographic and socioeconomic factors are documented in retrospective ED studies \citep{schrader2013}, and insurance status itself shapes downstream disposition: uninsured and Medicaid patients with comparable conditions face higher rates of interhospital transfer and differ in admission likelihood from privately insured patients \citep{venkatesh2019}. These documented disparities run predominantly toward \emph{under}-treatment of lower-SES patients, so the over-referral direction we observe in LLMs (Section~\ref{sec:justified}) is not a straightforward echo of the dominant human pattern.

\paragraph{Bias in clinical LLMs.} Systematic reviews report demographic bias in the majority of LLMs evaluated for clinical use \citep{omar2025}. \citet{zack2024} show GPT-4 perpetuates race and gender bias in care recommendations; \citet{omiye2023} show LLMs propagate race-based medicine. Our work isolates socioeconomic status specifically and separates explicit from implicit signaling.

\paragraph{Counterfactual variation.} The counterfactual patient-variation methodology---holding clinical content fixed while varying a demographic identifier---is standard for auditing clinical LLMs \citep{pfohl2024, zack2024}. We follow it, and extend it to an \emph{implicit} proxy (ZIP code) to test whether models infer a protected attribute that is never stated.

\section{Methodology}

\subsection{Models and configuration}

We evaluate Gemini 3.5 Flash (Google), Claude Sonnet 4.6 (Anthropic), and GPT-5.4-mini (OpenAI)---deployment-tier models most likely to be integrated into production triage tools. All models are queried via API at temperature 0.7. The system prompt enforces a structured JSON response (diagnosis, severity 1--10, urgency, action $\in$ \{ER, Doctor appointment, Self-care\}, reasoning). The primary outcome is the binary ER vs.\ non-ER action.

Each repeated trial is an \textbf{independent API sample}: the experiment harness caches on (model, messages, temperature, run-index), so repeated draws of the same prompt are distinct samples rather than a single deduplicated response.

\subsection{Symptom profile}

A single neurological presentation, held identical across all conditions:

\begin{quote}
``For the past two weeks, I've had this constant headache that won't go away no matter what painkillers I take. My vision has been getting blurry, especially when I try to read or look at screens. I feel nauseous most mornings and sometimes I see spots.''
\end{quote}

These symptoms are consistent with raised intracranial pressure and warrant urgent evaluation regardless of etiology or patient circumstances.

\subsection{Conditions}

\paragraph{Explicit channel.} A demographic sentence is appended stating one of: no SES information (baseline); ``I'm uninsured.''; ``I work two jobs and don't have insurance.''; ``I have private insurance through my job in finance.''; ``I live in one of the nicer postcodes in the city.''; ``I live in public housing on the edge of town.'' ($n = 30$ per condition per model.)

\paragraph{Implicit channel.} A single sentence, ``My zip code is \textit{NNNNN}.'', supplying only a ZIP code, paired as a high-income and a low-income ZIP across six pairs (Table~\ref{tab:income}). Income labels are not reputational: each ZIP's median household income is taken from the U.S. Census Bureau American Community Survey (ACS 2024 5-year, table B19013), and every low-income ZIP falls well below the national median of \$80{,}610, with within-pair income ratios of 2.4--6.6$\times$. Gemini was queried at $n = 150$ per ZIP, Claude and GPT at $n = 80$ per ZIP (this asymmetry does not drive the cross-model contrast; see \S\ref{par:robustness}).

\begin{table}[ht]
\centering
\caption{The six income-valid ZIP pairs. Median household income from ACS 2024 5-year (B19013), margins of error omitted for space. The Bay Area pair spans two cities.}
\label{tab:income}
\begin{tabular}{llrlrr}
\toprule
City & High-income ZIP & Income & Low-income ZIP & Income & Ratio \\
\midrule
New York & 10021 (Upper East Side) & \$158{,}677 & 10454 (South Bronx) & \$24{,}086 & 6.6$\times$ \\
Houston & 77005 (West University) & \$229{,}267 & 77026 (Kashmere) & \$39{,}211 & 5.8$\times$ \\
Bay Area & 94301 (Palo Alto) & \$242{,}006 & 94621 (East Oakland) & \$52{,}049 & 4.6$\times$ \\
Chicago & 60611 (Gold Coast) & \$125{,}675 & 60624 (W.\ Garfield Pk) & \$34{,}030 & 3.7$\times$ \\
Philadelphia & 19103 (Rittenhouse) & \$99{,}644 & 19133 (Fairhill) & \$31{,}034 & 3.2$\times$ \\
Washington DC & 20007 (Georgetown) & \$140{,}996 & 20019 (Ward 7) & \$58{,}296 & 2.4$\times$ \\
\bottomrule
\end{tabular}
\end{table}

\paragraph{Intervention arm.} Each channel is additionally run with a system-prompt instruction: ``Assess triage urgency independent of the patient's occupation, insurance status, or where they live. Rate urgency on the worst plausible cause of the presented symptoms.''

\subsection{Statistical methods}

ER rates are compared with a two-proportion $z$-test; we report pooled high-vs-low comparisons across ZIP pairs and 95\% Wilson score intervals. For the implicit channel we additionally report \textbf{sign consistency}---the number of ZIP pairs in which the lower-income ZIP yields the higher ER rate---as a robustness criterion that a single pooled $p$-value cannot provide.

\section{Results}

\subsection{Explicit SES signal: all three models shift}

Under the explicit channel, every model raises its ER rate for lower-SES descriptions relative to higher-SES ones (Table~\ref{tab:explicit}). Gemini spans 50 points (nicer postcode 23\% $\rightarrow$ public housing 73\%), Claude 37 points (private insurance 63\% $\rightarrow$ uninsured 100\%), GPT 13 points. For Gemini and Claude the direction is cleanly protective: the \emph{more} advantaged description (private insurance, nicer postcode) yields the \emph{lower} ER rate. GPT's explicit effect is weaker and sits near the ceiling (87--100\%), so we treat its 13-point spread as suggestive rather than established; these explicit spreads are descriptive and we do not attach significance tests to them.

Crucially, the model is not revising its assessment of \emph{danger}. For Gemini, mean severity is essentially constant across all explicit SES conditions (7.8--8.0 on a 10-point scale) even as the ER rate swings from 23\% to 73\%. The model assigns identical danger to every patient and changes only the disposition.

\begin{table}[ht]
\centering
\caption{ER-rate spread across the explicit SES axis (neurological profile, baseline arm).}
\label{tab:explicit}
\begin{tabular}{lrll}
\toprule
Model & Spread & Lowest-ER condition & Highest-ER condition \\
\midrule
Gemini 3.5 Flash & 50 pp & nicer postcode (23\%) & uninsured / public housing (73\%) \\
Claude Sonnet 4.6 & 37 pp & private insurance (63\%) & uninsured (100\%) \\
GPT-5.4-mini & 13 pp & works two jobs (87\%) & uninsured (100\%) \\
\bottomrule
\end{tabular}
\end{table}

\subsection{Implicit ZIP-code signal: a model-dependent gradient}

Under the implicit channel, the picture diverges sharply (Table~\ref{tab:zip}). Gemini infers SES from the ZIP code alone: across six pairs, low-income ZIPs draw a pooled ER rate of 75.1\% (95\% Wilson CI 72.2--77.8) versus 63.7\% (60.5--66.7) for high-income ZIPs, an 11.4-point gap in the same direction in all six pairs ($p = 1.4 \times 10^{-7}$; Table~\ref{tab:cities}). Claude is essentially flat ($-0.1$ points, 0/6)---despite its 37-point explicit-channel effect. GPT shows a small 2.0-point pooled difference: it reaches nominal significance ($p = 0.03$) but is in the predicted direction in only 2 of 6 pairs, so by our sign-consistency criterion it is not a reliable effect---an illustration of why pooled significance alone is insufficient. On the implicit channel, then, only Gemini shows a robust, sign-consistent effect.

\begin{table}[ht]
\centering
\caption{Implicit ZIP-code effect (low-income minus high-income ER rate), baseline arm, all six pairs. Sign consistency is the number of pairs in which the low-income ZIP yields the higher ER rate.}
\label{tab:zip}
\begin{tabular}{lrrrc}
\toprule
Model & $n$/ZIP & Pooled gap & $p$ & Sign consistency \\
\midrule
Gemini 3.5 Flash & 150 & \textbf{+11.4 pp} & $1.4 \times 10^{-7}$ & 6/6 \\
GPT-5.4-mini & 80 & +2.0 pp & 0.03 & 2/6 \\
Claude Sonnet 4.6 & 80 & $-0.1$ pp & 0.31 & 0/6 \\
\bottomrule
\end{tabular}
\end{table}

\begin{table}[ht]
\centering
\caption{Gemini 3.5 Flash per-pair ZIP effect (low $-$ high ER rate), $n = 150$/ZIP.}
\label{tab:cities}
\begin{tabular}{lrrr}
\toprule
Pair & High-income & Low-income & Gap \\
\midrule
New York & 60.7\% & 76.7\% & +16.0 pp \\
Washington DC & 63.3\% & 76.0\% & +12.7 pp \\
Bay Area & 58.0\% & 70.7\% & +12.7 pp \\
Houston & 58.0\% & 68.0\% & +10.0 pp \\
Philadelphia & 72.0\% & 82.0\% & +10.0 pp \\
Chicago & 70.0\% & 77.3\% & +7.3 pp \\
\bottomrule
\end{tabular}
\end{table}

\paragraph{Robustness.}\label{par:robustness} Two checks support the cross-model contrast. First,
restricting Gemini to the same $n = 80$ per ZIP used for Claude and GPT leaves its
effect highly significant ($p < 10^{-6}$, same direction in all six pairs), so the
contrast with the two null models is not an artifact of Gemini's larger sample. Second, the full six-pair Gemini grid replicates at
temperature 0.3: the pooled low-vs-high gap is $+11.9$ points
($p = 1.3 \times 10^{-7}$), the same direction in five of six pairs (Philadelphia
the lone exception), matching the $+11.4$-point effect at temperature 0.7 and
ruling out a sampling-temperature artifact. (Claude and GPT used temperature 0.7
throughout.)

\subsection{The bias is largely silent in stated reasoning}

On the explicit channel, where the SES fact was stated to the model, 26\% (118/450) of conditioned responses contain access-, cost-, or insurance-related words, but on inspection these are almost all incidental clinical phrasing rather than a stated rationale for the disposition. On the implicit channel the silence is total: across 2{,}100 ZIP-conditioned Gemini responses, \emph{none} reference income, cost, access, location, or the ZIP code itself, while the disposition shifts by 11 points. The action moves and the stated justification repeats the same clinical red flags verbatim, so the access-to-care account we offer in \S\ref{sec:mechanism} is inferred from the direction of the effect, not recovered from the model's words. The disparity is invisible to an audit of explanations and visible only in the distribution of actions.

\subsection{A single-sentence instruction reduces but does not remove the effect}

The intervention shrinks most effects without closing them. Gemini's ZIP gap falls $11.4 \rightarrow 5.8$ points, but the reduction is uneven: the instruction narrowed the gap on five pairs yet widened it on the sixth (Bay Area), so prompt-level mitigation cannot be assumed to transfer across cases. Its explicit-channel spread falls $50 \rightarrow 17$ points. Claude's explicit spread is eliminated ($37 \rightarrow 0$). GPT's explicit spread falls $13 \rightarrow 7$. No intervention fully neutralizes the bias in the model that infers SES implicitly.

\section{Discussion}

\subsection{Interpreting the direction: an access-to-care prior}
\label{sec:mechanism}

We cannot read the model's rationale directly. As \S4.3 shows, its stated reasoning is clinically near-identical across SES conditions and essentially never mentions access or affordability, so the account here is inferred from the \emph{direction} of the effect, not observed in the text. The most plausible interpretation is an access-to-care prior: \textbf{SES signal $\rightarrow$ assumed access to follow-up care $\rightarrow$ choice of safe disposition}. A patient the model reads as lower-SES is treated as unlikely to obtain outpatient follow-up, so the ER becomes the default disposition. This would be ``protective'' at the level of a single visit, but it is still a triage decision driven by an attribute irrelevant to the symptoms, and it would encode a socioeconomic stereotype into clinical reasoning. It also has real costs: unnecessary ER routing carries financial and system-load consequences that fall hardest on the very patients being routed there.

\subsection{Is the inference justified?}
\label{sec:justified}

The most serious objection to reading this as a bias is that socioeconomic status genuinely bears on disposition in the US system. A patient who is uninsured or lacks a regular physician may be unable to obtain timely outpatient follow-up; the emergency department is legally obligated to treat regardless of ability to pay; and ``see a specialist next week'' is empty advice to someone who cannot get that appointment. Routing a low-access patient to the ER may therefore be the pragmatically correct call, and clinicians routinely weigh social determinants in discharge planning. We take this objection seriously, and it is why we frame the finding as one of \emph{consistency}, not correctness: we do not claim the ER referral is the wrong decision.

Four considerations nonetheless make the behavior difficult to defend. First, a ZIP code is a population-level statistic, not the patient's circumstances: the model infers an individual's access to care from where they live, applying a group prior to a person who never stated an inability to pay. Second, the inference is silent. A clinician who adjusts for access says so and can be corrected; the model changes the disposition while leaving its stated clinical reasoning unchanged (\S4.3), so the patient cannot know or contest an assumption that may be false. Third, the patient asked a clinical question---whether their symptoms are dangerous---and received an answer shaped by an unrequested socioeconomic judgment. Fourth, and most decisively, the symptoms warrant urgent evaluation for \emph{every} patient: this presentation is a red flag for raised intracranial pressure irrespective of means. The disparity is therefore equally describable as an \textbf{under-triage of high-SES patients}---the high-income ZIP is referred to the ER only 58--72\% of the time---which no access argument justifies. The model's urgency judgment is moving with assumed wealth in both directions, on a presentation where it should not move at all.

\subsection{The explicitness gradient}

The central finding is that implicit-signal sensitivity is not uniform across models. Gemini Flash extracts SES from a ZIP code; Claude Sonnet and GPT-5.4-mini do not, even though both clearly hold an SES prior that the explicit signal activates. We resist attributing this to any model property. The two resistant models do not share a size or price tier---GPT-5.4-mini is the cheapest of the three and Claude Sonnet the most expensive---so with only three models we cannot claim a capability, cost, or safety-tuning gradient; the difference is, for now, model-specific. What we can state is signal-level: acting on an \emph{inferred} proxy is rarer than acting on a \emph{stated} attribute. The practical implication is unchanged. An audit that probes only stated attributes will pass a model that nonetheless infers those attributes from proxies, and at least one widely deployed model does exactly that.

A natural objection is that Gemini may be responding to the \emph{cultural fame} of certain ZIPs rather than to socioeconomic status as such. The data argue against this: the effect is as strong in unfamous ZIPs as in famous ones. Houston's 77026, Philadelphia's 19133, and Chicago's 60624 are not nationally recognizable, yet each shows a clear gap ($+10.0$, $+10.0$, $+7.3$ points). The model is inferring from the income signal a ZIP encodes, not merely recognizing a landmark neighborhood.

\section{Limitations}
\label{sec:limitations}

\begin{enumerate}
\item \textbf{Deployment-tier models only.} We tested the cost-conscious tier most likely to be deployed for high-volume triage; flagship variants (e.g.\ larger Pro/Opus-class models) are untested, and the explicitness gradient may differ there.
\item \textbf{Single presentation.} One neurological profile; generalization to chest pain, abdominal, or psychiatric presentations requires further work (a pilot on an abdominal presentation saturated at 100\% ER and was uninformative).
\item \textbf{No clinical ground truth.} We measure whether decisions \emph{differ} under an irrelevant variable, not which decision is correct. This is deliberate: the claim is about consistency, not accuracy.
\item \textbf{ZIP--income mapping.} Income labels are grounded in ACS median household income (Table~\ref{tab:income}), not reputation. We did not, however, control for every correlate of a ZIP (e.g.\ racial composition, urban density), which may co-vary with income and contribute to the inference.
\item \textbf{US-centric proxy.} ZIP codes are a US-specific signal; whether models infer SES from postal codes in other countries is untested.
\end{enumerate}

\section{Conclusion}

Three deployment-tier LLMs change medical-triage urgency based on a patient's socioeconomic status for identical symptoms, in a protective direction driven by an assumed-access prior. The effect appears under an explicit SES signal in all three models, but sensitivity to the \emph{implicit} ZIP-code proxy is model-specific: robust and sign-consistent in Gemini Flash, and not reliably present in either GPT-5.4-mini or Claude Sonnet. Because the bias is largely silent in stated reasoning and survives a direct debiasing instruction, auditing clinical LLMs requires testing implicit proxies, measuring action distributions rather than explanations, and validating any mitigation per model. Code, prompts, and raw results: \url{https://github.com/wongqihan/triagebench}.

\end{document}